# Dive into the heart of red giant stars to better understand our Galaxy

**Lagarde, N.**[1,2]

[1] *School of Physics and Astronomy Birmingham University, Birmingham, England*

[2] *Stellar Astrophysics Centre (SAC), Department of Physics and Astronomy, Aarhus University, Ny Munkegade 120, DK-8000, Aarhus C, Denmark*

**Abstract:** The availability of asteroseismic constrains for of large sample of stars observed with CoRoT and *Kepler* paves the way for statistical studies of the seismic properties of stellar populations, and becomes a powerful tool to better understand stellar structure and evolution. Here I present predictions of stellar models computed with the code STAREVOL including thermohaline mixing together with rotational mixing. I compare the theoretical predictions for the chemical properties of stars with recent spectroscopic of both field and cluster stars observations, and discuss the effects of both mechanisms on asteroseismic diagnostics, as well as on Galactic chemical evolution of helium-3.

## 1. INTRODUCTION

Stars are the building blocks of the Universe. Understanding their evolution is crucial to improve our knowledge from chemical properties of galaxies to the formation and evolution of planetary systems.

As shown by the initial mass function (e.g. [67]), low- and intermediate-mass stars form the dominant stellar component of our Galaxy and represent a very important astrophysical interest. After they leave the main sequence, these stars become red giants and undergo important changes of their structure and chemical composition. In these advanced phases, due to strong winds during the superwind phase, which leads to the emergence of planetary nebula, they contribute significantly to the enrichment of the interstellar medium and to the chemical evolution of galaxies. We are interested here specifically in their contribution to the evolution of $^3$He in the Galaxy.

Classical stellar evolution models of low- and intermediate-mass stars have difficulties to reproduce observations. In particular spectroscopic observations underline the existence of abundance anomalies at the surface of many red giant stars, and provide compelling evidence of non-canonical mixing process that occurs when low-mass stars reach the so-called bump in the luminosity function on the red giant branch (RGB). At that phase, the surface carbon isotopic ratio drops, together with the abundances of lithium and carbon, while nitrogen increases slightly (e.g. [41], [44], [70]). Standard models do not explain these changes on stellar surface abundances along the RGB (see Salaris M. contribution in this volume).

Different transport processes have been proposed in the literature (e.g. [12], [30], [74], [63], [21], [22]) to explain these abundance anomalies observed in red giant stars. More particularly, rotation-induced mixing has been investigated as a possible source of mixing on the RGB. It has been found however that rotation-induced mixing has an impact on stellar structure and on the chemical surface abundances during the main sequence, but it does not explain abundance anomalies observed in low-mass red giants ([61],[62],[63]). I present in part 2 this transport process together with thermohaline instability, which has been proposed by [21] to be a fundamental physical process in low-mass red giant stars. In part 3, I discuss effects of these two transport processes on surface abundances, and their efficiency with the initial stellar mass and metallicity.

In recent years, a large number of asteroseismic observations have been obtained for different kinds of stars. The comparison between models including a detailed description of transport processes in stellar interiors and asteroseismic constraints from the space missions CoRoT and *Kepler*, opens a new promising path for our understanding of stars. In part 4, I present how asteroseismology can be a very useful tool added to spectroscopy to better constrain stellar evolution.

The predictions of these new stellar models including both thermohaline instability and rotation-induced mixing show that these two transport processes have an impact on the chemical composition of the material ejected by the star in the interstellar medium. I also present, in part 5, their consequences on the evolution of $^3$He in our Galaxy, and how they help us to resolve the long-standing "$^3$He problem".

## 2. TRANSPORT PROCESSES INSIDE RED GIANT STARS

Numerous spectroscopic observations (e.g. [72], [26], [68], [44], [78], [79], [56]) provide compelling evidence for non-canonical processes that modify the surface abundances of low- and intermediate-mass stars, which are not predicted by standard stellar theory. In the following, I present the two most likely transport processes, which have been proposed in the literature to explain abundance anomalies in giants.

### *2.1. Rotation-induced mixing*

Rotation has been investigated as a possible source of mixing in RGB stars by several authors (e.g. [74], [18], [31], [63], [15]). In this section, we will first briefly summarize the current state-of-the-art description of rotation-induced mixing in stars, and its effects on abundances profiles and nucleosynthesis of low- and intermediate-mass stars. All models presented in this paper have been computed with the stellar evolution code STAREVOL ([69], [62], [28], [49]).

*2.1.1. Physics*

For the treatment of rotation-induced mixing, we use the complete formalism developed by [85] and [52] (for a description of the implementation in STAREVOL, see [62], [63], [28]). The transport of angular momentum in stellar radiative layers obeys an advection/diffusion equation (1), where we can identify the terms A, B, and C describing respectively the stellar contraction and expansion; advection of angular momentum by meridional circulation; and the diffusion effect of shear-induced turbulence.

$$\underbrace{\rho \frac{d(r^2\Omega)}{dt}}_{A} = \underbrace{\frac{1}{5r^2}\frac{\partial}{\partial r}\left(\rho r^4 \Omega U_r\right)}_{B} + \underbrace{\frac{1}{r^2}\frac{\partial}{\partial r}\left(r^4 \rho \nu_v \frac{\partial \Omega}{\partial r}\right)}_{C} \qquad (1)$$

where r is the stellar radius, $\rho$ the density, $\nu_v$ the vertical component of the turbulent viscosity, and $\Omega$ the angular velocity, and $U_r$ the vertical component of meridional circulation.

The transport of chemicals resulting from meridional circulation and both horizontal and vertical turbulence is computed as a diffusive process throughout evolution, and follows (2)

$$\frac{dc_i}{dt} = \dot{c}_i + \frac{1}{\rho r^2}\frac{\partial}{\partial r}\left(r^2 \rho D_{tot} \frac{\partial c_i}{\partial r}\right) \quad (2)$$

where $c_i$ the concentration of a chemical species i, and $\dot{c}_i$ represents their variations due to nuclear reactions. The total diffusion coefficient $D_{tot}$ for chemicals can be written as the sum of two diffusion coefficients (3)

$$D_{tot} = D_{eff} + D_v \quad (3)$$

with $D_{eff}$ the effective diffusion coefficient by [85] and [14], $D_{eff} = \frac{|rU(r)|^2}{30 D_h}$, depending on the horizontal diffusion coefficient by [85]; and $D_v$ the vertical turbulence diffusion coefficient ([75]).

*2.1.2. Effects in low and intermediate-mass stars*

Rotation-induced mixing on the main sequence modifies the internal and surface chemical abundances as extensively tested in previous papers. Reference [49] (see their references) have also largely investigated stellar evolution in massive stars. It accounts nicely for the behavior of lithium and beryllium at the surface of Population I main-sequence and subgiant low-mass stars (see [76], [77], [20], [62], [64], [71], [22]).

Rotation has also an impact on the internal abundance profiles of heavier chamicals involved in hydrogen-burning at higher temperature than the fragile Li and Be. Fig. 1 represents the abundances profiles at the end of the main sequence for a 1.5 Msun model computed with (at different initial velocity) and without rotation (top left panel). In the rotating model, the abundance gradients are smoothed out compared to the standard case: $^3$He, $^{13}$C, $^{14}$N, and $^{17}$O diffuse outwards, while $^{12}$C and $^{18}$O diffuse inwards. However, rotation-induced mixing is not efficient enough to noticeably change the surface abundances of these elements on the main sequence for the 1.5 $M_\odot$ model, although it sets the scene for abundance variations in latter evolution phases. In particular the surface abundance variations during the first dredge-up are slightly strengthened when rotation-induced mixing is accounted for. For example, more $^3$He is brought into the stellar envelope, and the post dredge-up $^{12}$C/$^{13}$C is lower than in the non-rotating case.

As shown by [63] who studies the effects if rotation-induced mixing in the RGB for low-mass stars, the total diffusion coefficient of rotation is too low to reproduce variations of surface abundances on the first ascent giant branch as requested by spectroscopic observations of RGB brighter than the bump. Another transport process should be invoked to complement the effects of rotation on the red giant branch.

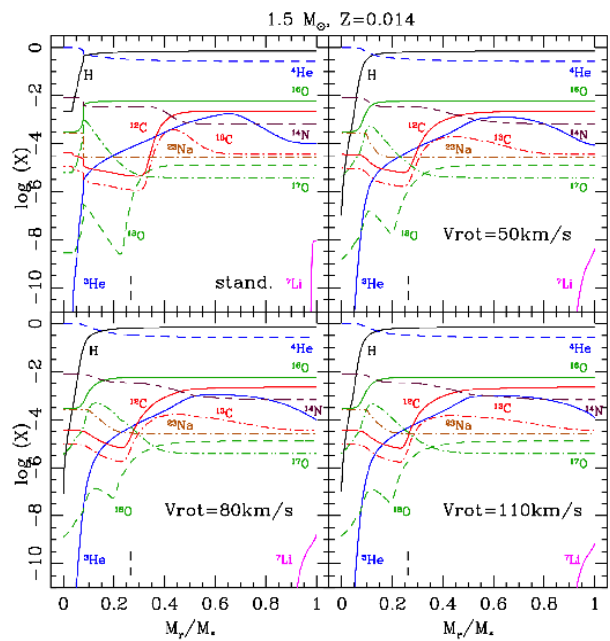

Figure 1. Chemical structure at the turnoff of 1.5 $M_\odot$ star computed for three different initial rotation velocities. The abundances are given in mass fraction and are multiplied by 100 for $^3$He, $^{12}$C, and $^{14}$N, by 2500 for $^{13}$C, by 50, 900, and $5.10^4$ for $^{16}$O, $^{17}$O, and $^{18}$O respectively, and by 1500 for $^{23}$Na. The vertical arrows show, in all cases, the maximum depth reached by the convective envelope at its maximum extent during the first dredge-up. *Figure from [51].*

## 2.2. Thermohaline instability

### 2.2.1. Physics

Thermohaline mixing has recently been identified as the mechanism that governs the photospheric composition of low-mass bright giant stars ([21]). In such stars, this double-diffusive instability is induced by the molecular weight inversion created by the $^3$He($^3$He,2p)$^4$He reaction in the external wing of the hydrogen-burning shell ([34], [35]). Indeed this peculiar reaction converts two particles into three and thus decreases the mean molecular weight, as already pointed out by [81] although in a different stellar context. The thermohaline instability is expected to set in after the first dredge-up when the star reaches the RGB luminosity bump. In terms of stellar structure, the RGB bump corresponds to the moment when the hydrogen-burning shell encounters the chemical discontinuity created inside the star by the convective envelope at its maximum extent during the first dredge-up. When the source shell (which provides the stellar luminosity on the RGB) reaches the border of the H-rich previously mixed zone, the corresponding decrease in molecular weight of the H-burning layers induces a drop in the total stellar luminosity, thereby creating a bump in the luminosity function since stars spend a relatively long time at this location (i.e., [37], [17], [19]). Afterwards, H-burning occurs in a region of uniform composition, leading to a molecular weight inversion due to $^3$He burning and thus enabling the thermohaline instability to set in.

Thermohaline instability occurs in a stable stratification that satisfies the Ledoux criterion for convection instability:

$$\nabla_{ad} - \nabla + \left(\frac{\varphi}{\delta}\right)\nabla_\mu > 0 \qquad (4)$$

and where the molecular weight decreases with depth:

$$\nabla_\mu = \frac{d\ln\mu}{d\ln P} < 0 \qquad (5)$$

with the classical notation for $\nabla = \partial\ln T / \partial\ln P$; $\varphi = (\partial\ln\rho/\partial\ln\mu)_{P,T}$; and $\delta = -(\partial\ln\rho/\partial\ln T)_{P,\mu}$; $\nabla_\mu$ and $\nabla_{ad}$ are respectively the molecular weight gradient and adiabatic gradient. To compute thermohaline instability, we use the prescription advocated by [21] based on [78] arguments for the aspect ratio α (length/width) of salt fingers as supported by laboratory experiments by [46], including [47] expression for non-perfect gas:

$$D_{th} = C_t K \left(\frac{\varphi}{\delta}\right)\frac{-\nabla_\mu}{\nabla_{ad} - \nabla}, \quad \text{for} \quad \nabla_\mu < 0 \qquad (6)$$

with K the thermal diffusivity, $C_t = \frac{8}{3}\pi^2\alpha^2$; and α = 5.

*2.2.2. How does the thermohaline instability occur in red giant stars ?*

Fig. 2 represents the abundance profiles of selected elements and the mean molecular weight gradient $\nabla_\mu = d\ln\mu/d\ln P$, before (left panel) and after (two right panels) the bump in the 1.5 $M_\odot$, at solar metallicity star as a function of the relative mass[1]. During the first dredge-up episode, the convective envelope homogenizes down to very deep regions, and builds a step in the gradient of molecular weight at its maximum penetration, which corresponds to the external peak (δM=0.85 and δM=0.2 respectively on left panel and middle/right panels of Fig. 2) in the profile of molecular weight gradient. The internal peak (at δM~0.08) corresponds to the region where H is efficiency depleted by nuclear reactions in the hydrogen burning shell (HBS). Before the bump, the gradient of mean molecular weight is positive in the whole radiative region, and the thermohaline instability cannot set in (see (6)). The profiles of chemical species are thus identical to those obtained in a standard model.

At the bump, when the HBS passes through the μ-discontinuity left behind the first dredge-up, H burns in a homogeneous region. The $^3$He($^3$He,2p)$^4$He reaction lowers the molecular weight in the external wing of the HBS and then $\nabla_\mu$ becomes negative(magenta bold dashed line of middle and right panels, see [22] for more details). As shown by middle and right panels, the thermohaline instability develops between the $^3$He-burning region and the convective envelope. In the right panel, the thermohaline instability "connects" these two regions, which leads to abundances changes (see abundances profiles of $^3$He, $^{12}$C, $^{13}$C, and $^7$Li on Fig. 2) at the stellar surface. In deeper radiative regions, $\nabla_\mu$ remains positive (bold back line) and no thermohaline mixing occurs.

---

[1] δM=($M_r$-$M_{core}$)/($M_{env}$-$M_{core}$) the relative mass is defined as ranging from 0 to 1 between the bottom of the hydrogen burning shell and the base of the convective envelope.

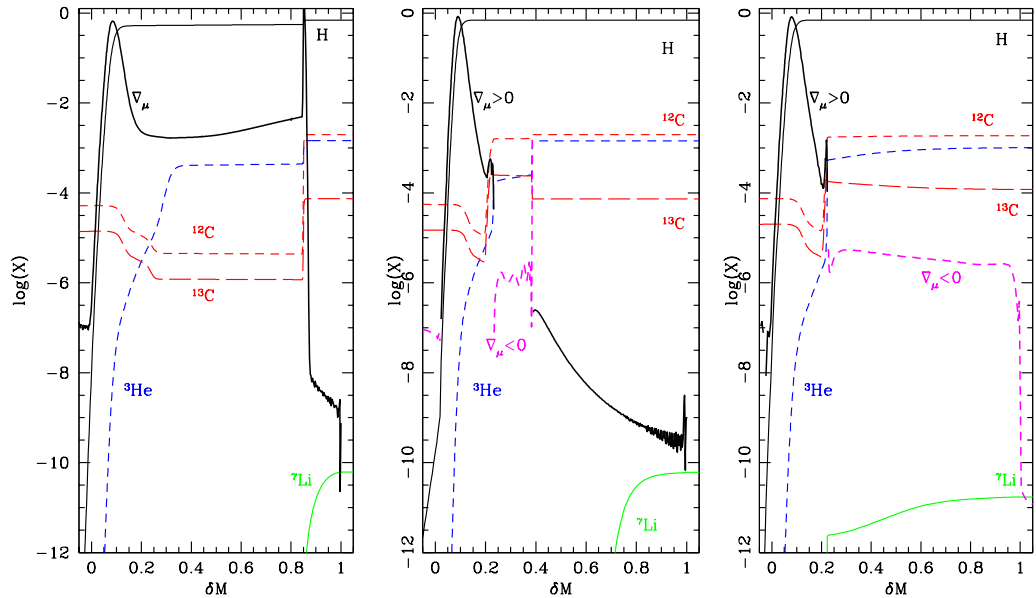

Figure 2: Profiles of the logarithm of the mass fraction of H, $^3$He, $^{12}$C, $^{13}$C, $^7$Li and of the mean molecular weight gradient $\nabla_\mu = d\ln\mu / d\ln P$ (bold line, full when $\nabla_\mu > 0$ dashed otherwise) as a function of the reduced mass (see text) inside a 1.0 M$_\odot$, solar metallicity star. Left to the right panels correspond to models including thermohaline mixing at various luminosities along the RGB. The left panel is located just before the bump, and others correspond to models after the bump luminosity

As a consequence, the surface abundance of $^3$He, $^{12}$C, $^{13}$C (see Fig. 1), and $^{14}$N are already modified soon after the onset of thermohaline mixing. However, as shown by [21], [22] the surface abundance of $^{16}$O, and $^{23}$Na remain constant because the thermohaline mixing does not extend down to the very deep region where full CNO-burning operates at equilibrium.

### 3. COMPARISON BETWEEN MODELS AND SPECTROSCOPIC OBSERVATIONS…

Let us now compare models predictions including rotation-induced mixing and thermohaline instability, with spectroscopic observations of carbon isotopic ratio in open clusters, and in field stars.

### 3.1. …all along the evolution

Fig. 3 presents observations of $^{12}C/^{13}C$ ratio in sub giant, giant, and clump stars in the open cluster M67 by [42]. The data are compared to the predictions of 1.25 $M_\odot$ models following standard prediction (left panel), and including thermohaline instability and rotation induced mixing at three different initial velocities. We see that standard prediction does not reproduce low value of carbon isotopic ratio in evolved stars. On the other hand, the theoretical predictions of the models including the two transport processes explain very well the observational behavior all along the evolutionary sequence. While the dispersion for stars that have not yet reached the RGB bump (i.e. before $\log(L/L_\odot)\sim1.8$) is explained only with the dispersion of initial rotation velocity, thermohaline mixing accounts well for the surface abundances of stars more evolved.

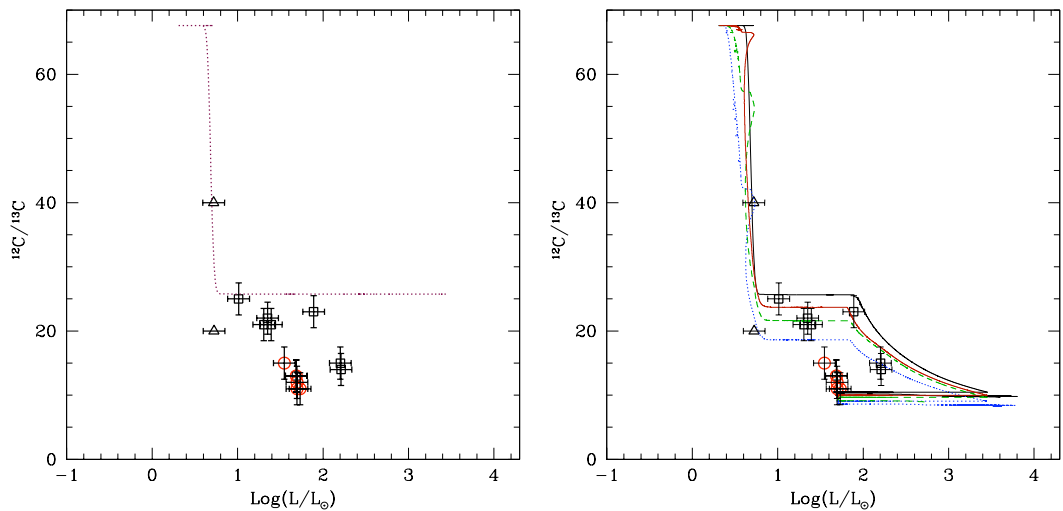

Figure 3. Evolution of the surface $^{12}C/^{13}C$ value as a function of stellar luminosity for the 1.25$M_\odot$ models at solar metallicity following standard prediction (left panel); and including thermohaline instability and rotation-induced mixing (right panel, for initial rotation velocities of 50, 80, and 110 km.s$^{-1}$ shown as solid red, dashed green, dotted blue lines respectively; the non-rotating case is also shown with the black solid line). Observations along the evolutionary sequence of the open cluster M67 are from [42]. The black triangle is for subgiant star for which only a lower value could be obtained, while black squares and red circles correspond respectively to RGB and clump stars. *Figures from [22]*

### 3.2. Efficiency as a function of stellar mass

We would like, now, to underline, the efficiency of transport processes as a function of initial stellar mass. For that, we compare in Fig. 4 the predictions of our models over 1-4 $M_\odot$ range at solar metallicity with carbon isotopic ratio in open clusters of different turnoff masses. For details on the open clusters presented here please refer to [22].

Models predictions are shown both at the RGB tip (black line) and after the 2DUP (dashed lines) for different assumptions. Dotted lines correspond to standard models predictions. Solid lines correspond to models computed with thermohaline mixing only. They reproduced very well the $^{12}C/^{13}C$ behavior for evolved stars with masses lower than ~1.7$M_\odot$. Indeed, as we discussed in part 2, thermohaline mixing is induced by $^{3}He(^{3}He,2p)^{4}He$ reaction, and so its efficiency depends on the amount of $^{3}He$ available to power thermohaline mixing. On the main sequence, low-mass stars burn hydrogen mainly through the pp-chains rather than CNO cycle in intermediate-mass. Consequently, a large production of $^{3}He$ in low-mass stars. In this mass range rotation-induced mixing leads only to slightly lower values as shown by dashed lines for initial velocity of 110 km.s$^{-1}$.

Intermediate-mass stars, with masses between ~1.7$M_\odot$ and 2.2 $M_\odot$, burn hydrogen mainly through the CNO cycle, thus thermohaline mixing is less efficient than in previous mass range. Here, both mecanisms play an equivalent role to change surface abundances.

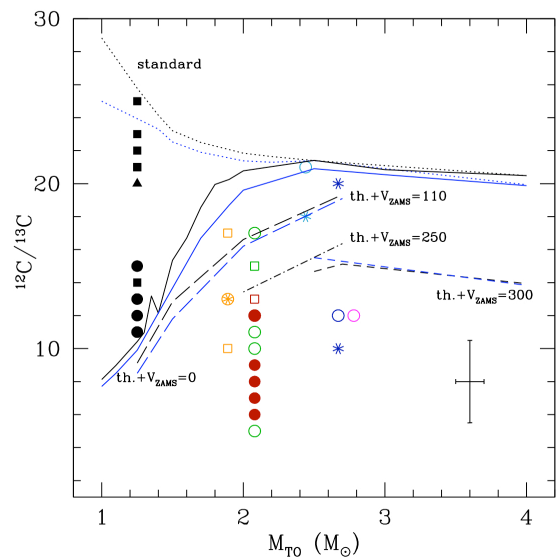

Figure 4. Observations of $^{12}C/^{13}C$ in evolved stars of Galactic open clusters by [71] (open symbols), [41], [42], and [56] as a function of the turnoff mass of the corresponding host cluster that can be identified thanks to the colours of the symbols. Squares, circles, and asteriscs are for RGB, clump, and early-AGB stars respectively, while diamonds are for stars from [41] sample with doubtful evolutionary status; triangles are for lower limits. A typical error bar is indicated. Theoretical predictions are shown at the tip of the RGB and after completion of the second dredge up (black and blue lines respectively). Standard models (no thermohaline nor rotation induced mixing) are shown as dotted lines, models with thermohaline mixing only (V$_{ZAMS}$ = 0) as solid lines, and models with thermohaline and rotation-induced mixing for different initial rotation velocities as indicated as long-dashed, dot-dashed, and dashed lines. *Figure from [22]*

And finally, for more massive stars, thermohaline mixing plays no role, because these stars ignite central He burning before reaching the RGB bump. As results, thermohaline instability does not occur in these stars, only rotation-induced mixing has an impact on surface abundances.

### 3.3. Efficiency as a function of metallicity

Fig. 5 represents the surface $^{12}C/^{13}C$ as function of initial stellar mass for models following different assumptions at two metallicities (solar and [Fe/H]=-0.56 in black and blue lines respectively). Observations of $^{12}C/^{13}C$ in field stars in the metallicity range between [Fe/H]=-0.52 and [Fe/H]= 0.17, are from [79]. Theoretical and observational behaviours are in agreement for this metallicity range. In addition, at a given stellar mass, the efficiency of transport processes to reduce $^{12}C/^{13}C$ at the stellar surface increases when the metallicity decreases. The effect of metallicity on the efficiency of both transport processes will be discussed in the forthcoming paper (Lagarde & Charbonnel 2013 in prep)

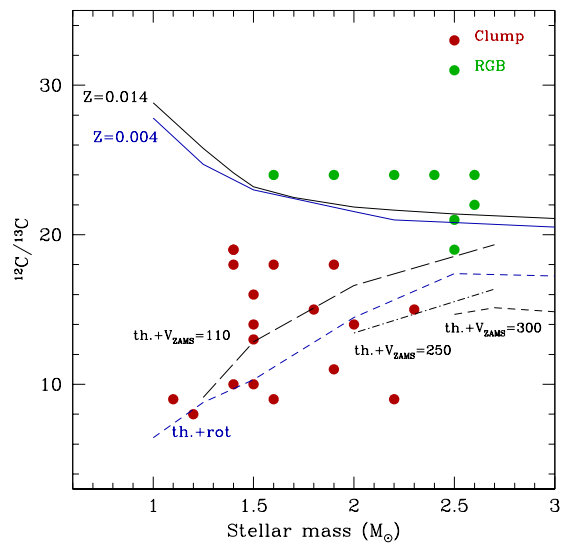

Figure 5. Observations of $^{12}C/^{13}C$ in evolved field stars (green and red circles respectively represent RGB and clump stars) by [79] as a function of stellar mass. Theoretical predictions are shown at the end of the first dredge-up (solid lines) and after completion of the second dredge up (dashed lines), and for two metallicities: [Fe/H]=0 and [Fe/H]=-0.56 shown by black and blue lines respectively. The initial rotation velocity of low metallicity model took equal to 45% of critical velocity.

*3.4 Conclusions*

By comparison of our models with chemical abundance determinations in open clusters and in field stars, I underlined the following conclusions:

- Thermohaline instability occurs when low- and intermediate-mass stars reach the so-called bump in the luminosity function on the Red Giant Branch (RGB). At this phase, the double diffusive instability is induced by $^3$He($^3$He,2p)$^4$He reaction that creates an inversion of mean molecular weight. I showed that its efficiency increases with the decrease of the initial stellar mass, and metallicity.
- During this phase thermohaline mixing induces the changes of surface abundances of $^3$He, $^7$Li, C and N for stars brighter than the bump luminosity. Our model predictions are compared to observational data for lithium, $^{12}$C/$^{13}$C in Galactic open clusters and in field stars with well defined evolutionary status. Thermohaline mixing simultaneously reproduce the observed behaviour of $^{12}$C/$^{13}$C; [N/C], and lithium (presented in [22]) in low-mass stars that are more luminous than the RGB bump.
- Rotation-induced mixing modifies the internal chemical structure of main sequence stars. The dispersion of chemicals surface abundances in evolved stars is very well explained when accounting for a dispersion in the initial values of the stellar rotational velocity as observed in young open clusters.

## 4. ASTEROSEISMOLOGY

Contrary to spectroscopic observations that give information limited to the stellar surface layers, asteroseismology allows us to constrain stellar interiors by the study of stellar pulsation modes. Thousands of evolved stars (subgiant, giant and clump stars) have been already observed by CoRoT ([1]) and *Kepler* ([11]). Determination of individual frequencies represents an excellent opportunity to deduce from asteroseismology stellar mass and radius (see (7)), as well as distance and age (see [16] and their references).

$$\left(\frac{R}{R_{sun}}\right) = \left(\frac{\nu_{max}}{\nu_{max,sun}}\right)\left(\frac{\Delta\nu}{\Delta\nu_{sun}}\right)^{-2}\left(\frac{T_{eff}}{T_{eff,sun}}\right)^{0.5}$$

$$\left(\frac{M}{M_{sun}}\right) = \left(\frac{\nu_{max}}{\nu_{max,sun}}\right)^{3}\left(\frac{\Delta\nu}{\Delta\nu_{sun}}\right)^{-4}\left(\frac{T_{eff}}{T_{eff,sun}}\right)^{1.5}$$

(7)

In addition, asteroseismology brings key information on the stellar structure with different acoustic radius (total, at the base of convective envelope or the base of the helium second ionization region), and with the period spacing of gravity modes. Fig. 7 presents the period spacing of gravity modes $\Delta\Pi(l=1)$ for standard model of 2.0 M$_\odot$ at solar metallicity. As proposed by [9], [58], this quantity allows us to distinguish two stars that have the same luminosity, one being on the RGB and the other on the clump undergoing the central He-burning. Indeed the stellar structure, and the presence of convective core affects the domain where the g-modes are trapped, and then the value of $\Delta\Pi(l=1)$. This quantity is theoretically expected to be larger in clump stars than in RGB stars ([25]).

Moreover, thanks to asteroseismology, we can now determine the internal rotation profile of giant stars ([29], [7], [59]), and then in the future better test models of transport of angular momentum ([33], [13]).

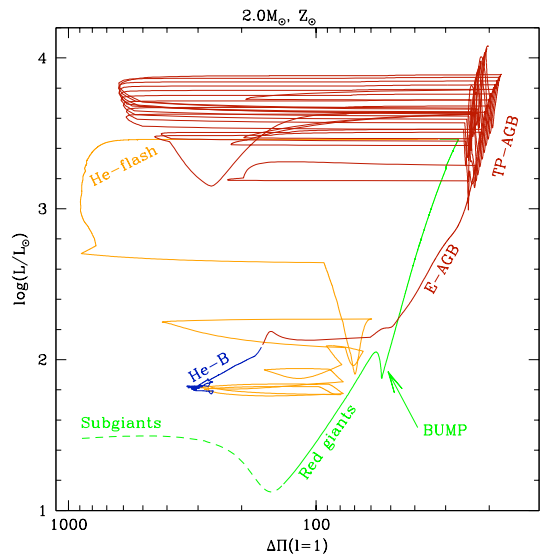

Figure 6. Stellar luminosity as a function of the asymptotic period spacing of g-modes for the standard 2.0 M_model at solar metallicity. Evolutionary phases are color-labeled: subgiant (green dashed), red giant (green solid), helium-flash episode (orange), helium-burning (blue), and asymptotic giant branch (red). *Figure from [49]*

To follow this new open way, we provide the relevant classical stellar parameters together with the asteroseismic properties of low- and intermediate-mass stars including effects of thermohaline instability and rotation-induced mixing in a grid of stellar models computed at different metallicities ([49]).

## 5. GALACTIC CHEMICAL EVOLUTION OF HELIUM-3

### 5.1 "$^3$He problem"

The classical theory of stellar evolution predicts a very simple Galactic destiny to $^3$He, dominated by the production of this isotope during Standard Big Bang Nucleosynthesis (SBBN) and in stars with initial mass lower than ~ 3 $M_\odot$ (see [45], [47]).
In these objects, $^3$He is produced first through D-processing on the pre-main sequence and then through the pp-chain on the main sequence. This fresh $^3$He is then engulfed in the stellar envelope during the so-called first dredge-up ([45]), and according to classical stellar modelling it survives the following evolution phase before being released in the interstellar matter through stellar wind and planetary nebula ejection ([65], [82], [27], [83], [36]).

As a consequence, one expects with time a large increase of $^3$He in the Galaxy (see Fig. 7, red solid line on left panel) with respect to its primordial abundance (see e.g. [84]); this latest quantity is well constrained thanks to accurate determination of the baryon density of the Universe by recent cosmic microwave background experiments, most particularly from WMAP ([10], [73]), which has led to an unprecedented precision on the yields of SBBN. Galactic HII regions should in particular be highly enriched in $^3$He since their matter content chronicles the result of billion years of chemical evolution since the Milky Way formation. Additionally the $^3$He/H abundance ratio is expected to be higher in the central regions of the Galaxy where there has been more substantial stellar processing than in the solar neighborhood (see Fig. 7, red solid line on right panel).
However, the $^3$He abundance in HII regions sampling a large volume of the Galactic disk is found to be very homogeneous ([66], [2], [3], [4], [5], [6]), similar to that of the Sun at the epoch of its formation (for references see [40]), and only slightly higher than the WMAP+SBBN primordial abundance. No observational evidence is thus found for strong enrichment of this element along Galactic history contrary to expectations of all chemical evolution models that take into account $^3$He yields from classical stellar models (e.g. [38], [60], [80]). This is the well-known **"$^3$He problem"**.

### *5.2 Galactic evolution of helium-3*

In previous parts, I described the effects of thermohaline instability and rotation-induced mixing on the surface abundances of low- and intermediate-mass stars. I also showed that these processes change the surface abundances in giant stars. Reference [45], showed that thermohaline mixing has an impact on the surface abundance of $^3$He in low-mass stars, and then on the quantity of $^3$He ejected in the interstellar medium by these stars. For Low-mass stars (M<2-2.2 $M_\odot$) that produce large quantities of this light element through the pp chains on the main sequence, thermohaline mixing on both the RGB and the TP-AGB is dominant in reducing the final $^3$He yield. These stars remain net producers of $^3$He however, although their contribution to the Galactic evolution of this light element is strongly reduced compared to the standard framework. For intermediate-mass stars with masses between 2-2.2 and 3-4 $M_\odot$, thermohaline mixing leads also to modest $^3$He depletion during the TP-AGB phase, associated with lithium production.

In Fig. 7, I present predictions of Galactic Chemical Evolution (GCE) models (more details of the galactic chemical evolution code in [54], [23], [55], and [24]), including thermohaline instability and rotation-induced mixing (dashed blue line). GCE models reproduce the observations of $^3$He in the proto-solar cloud (PSC), local interstellar medium (LISM) and HII regions, while $^3$He is overproduced on a Galactic scale with standard models. We concluded that thermohaline mixing is the only physical mechanism known so far able to solve the so-called "$^3$He problem" plaguing in the literature since many years.

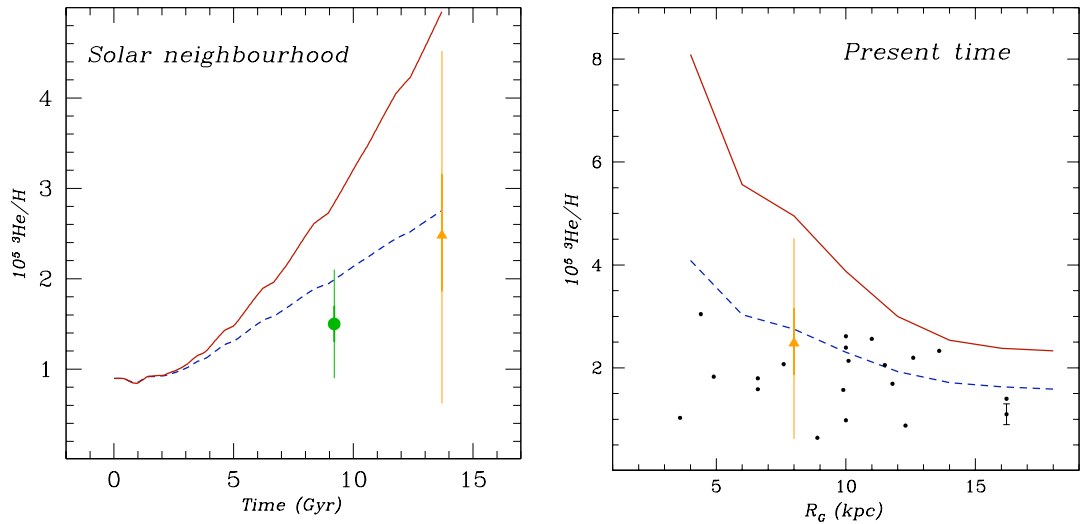

Figure 7. Left panel: evolution of $^3$He/H with time in the solar neighborhood. Data for the proto solar cloud (green filled circle) and local interstellar medium (LISM, orange filled triangle) are from [39] and [43], respectively. 1-σ and 3-σ error bars are shown with thick and thin lines, respectively. Right panel: radial distribution of $^3$He/H at the present time. The dots are HII regions data from [6] (error bars are shown only for S209; see text for discussion). The triangle at $R_G$ = 8 kpc represents LISM data from [43]. The predictions from standard stellar evolution and from models including thermohaline instability and rotation are shown in both panels by the red full, blue dashed lines respectively. *Figure from [50]*

## 6. CONCLUSIONS

I showed that that thermohaline mixing is the dominating chemical transport process in low-mass red giant stars (M ≤ 2.0 M$_\odot$), which governs their photospheric composition, while rotation-induced mixing with different initial velocities, explains surface abundances in more massive stars (M>2.0 M$_\odot$). Including in the galactic chemical evolution code, new stellar yields of 3He taking into account effects of thermohaline instability and rotation-induced mixing, we showed that these new stellar models provide a solution to the long standing "$^3$He problem" on Galactic scale.

Furthermore, CoRoT and *Kepler* missions have detected solar-like oscillations in thousands of red giant stars ([32], [8]), which will add valuable and independent constraints on current stellar models. These observations are sampling different regions of the Galaxy, and promise to improve our understanding on the Milky Way's constituents. To exploit all the potential of asteroseismic data from CoRoT and *Kepler* missions, it would be crucial to combine them with spectroscopic constraints, which should be available in the future with the large spectroscopic survey such as SDSS-APOGEE, and GAIA mission (~ 1 billion stars essentially in the Milky Way). Meanwhile, we will compare our model predictions with some stars for which spectroscopic studies have been done, to better constrain theoretical stellar evolution and effects of transport processes (see [57]).

### ACKNOWLEDGMENT

I acknowledge financial support from the Swiss National Fund, and funding for the Stellar Astrophysics Centre which is provided by The Danish National Research Foundation (Grant agreement no.:DNRF106).